\documentclass[aoas,MSNbibl,nameyear,dvips]{arximspdf}
\usepackage{graphicx}
\doi{10.1214/14-AOAS728} % kopijuoti is laisko
\volume{8}
\issue{4}
\pubyear{2014}
\firstpage{1922}
\lastpage{1946}
\docsubty{FLA}
\relateddois{T11}{Discussed in \relateddoi{d}{10.1214/14-AOAS728ED},
\relateddoi{d}{10.1214/14-AOAS728A},
\relateddoi{d}{10.1214/14-AOAS728B} and
\relateddoi{d}{10.1214/14-AOAS728C};
rejoinder at \relateddoi{r}{10.1214/14-AOAS728R}.}

\makeatletter
\def\cal{\mathcal}

\makeatother

\begin{document}
\begin{frontmatter}

\title{Spatial accessibility of pediatric primary healthcare:
Measurement and inference\thanksref{T11,T1}}
\runtitle{Accessibility of pediatric primary healthcare}
\thankstext{T1}{Supported in part by the NSF Grant CMMI-0954283 and a seed grant awarded by the
Healthcare System Institute and Children's Healthcare of Atlanta.}

\begin{aug}
\author[A]{\fnms{Mallory}~\snm{Nobles}\ead[label=e1]{mallory.nobles@gmail.com}},
\author[A]{\fnms{Nicoleta}~\snm{Serban}\corref{}\ead[label=e2]{nserban@isye.gatech.edu}}
\and
\author[A]{\fnms{Julie}~\snm{Swann}\thanksref{T2}\ead[label=e3]{swann@isye.gatech.edu}}
\thankstext{T2}{Supported by the Harold R. and Mary Anne Nash Junior
Faculty Endowment Fund.}
\runauthor{M. Nobles, N. Serban and J. Swann}
\affiliation{Georgia Institute of Technology}
\address[A]{H. Milton Stewart School of Industrial\\
\quad and Systems Engineering\\
Georgia Institute of Technology
755 Ferst Dr. NW\\
Atlanta, Georgia 30332\\
USA\\
\printead{e1}\\
\phantom{E-mail:\ }\printead*{e2}\\
\phantom{E-mail:\ }\printead*{e3}} %adresu isvedimo komanda gale!
\end{aug}

% HISTORY:
\received{\smonth{1} \syear{2013}}
\revised{\smonth{1} \syear{2014}}

% ABSTRACT
%
\begin{abstract}
Although improving financial access is in the spotlight of the current
U.S. health policy agenda, this alone does not address universal and
comprehensive healthcare. Affordability is one barrier to healthcare,
but others such as availability and accessibility, together defined as
\textit{spatial accessibility}, are equally important. In this paper, we
develop a measurement and modeling framework that can be used to infer
the impact of policy changes on disparities in spatial accessibility
within and across different population groups. The underlying model for
measuring spatial accessibility is optimization-based and accounts for
constraints in the healthcare delivery system. The measurement method
is complemented by statistical modeling and inference on the impact of
various potential contributing factors to disparities in spatial
accessibility. The emphasis of this study is on children's
accessibility to primary care pediatricians, piloted for the state of
Georgia. We focus on disparities in accessibility between and within
two populations: children insured by Medicaid and other children. We
find that disparities in spatial accessibility to pediatric primary
care in Georgia are significant, and resistant to many policy
interventions, suggesting the need for major changes to the structure
of Georgia's pediatric healthcare provider network.
\end{abstract}

% KEYWORDS
\end{frontmatter}

%s1 #&#
\section{Introduction}\label{sec1}

%Background and Motivation
Starting in 2014, U.S. health policy will undergo a significant, albeit
strongly debated, transformation through the Affordable Care Act. This
bill, like many of the recent initiatives in health policy, focuses on
improving financial access to health coverage for all Americans,
including the 50 million uninsured. While health services can indeed be
prohibitively expensive, many other factors can also limit patients'
ability to participate in the healthcare system. In addition to
affordability, there are at least four other barriers to healthcare
access: availability, accessibility, acceptability and accommodation
[\citet{PenTho81}]. For policy makers to design effective
strategies for improving all patients' access to healthcare, it is
important that they understand and consider each of these dimensions.
Affordability or financial access alone does not guarantee universal
and comprehensive access to healthcare.

% What is the paper about?
In this paper our emphasis is on two dimensions of access, \textit{availability}, or the number of local service sites from which a
patient can choose, and \textit{accessibility}, which considers the time
and distance impediments between patient locations and service sites.
These two potential barriers to healthcare are driven by differences in
geographic access and are referred to as \textit{spatial
accessibility} in
the existing literature [\citet{JosPhi84}; Guagliardo et al.
(\citeyear{Guaetal04}); \citet{McGHum09}]. Specifically, we develop a
measurement and modeling framework that can be used to infer the impact
of policy changes on the equity of spatial accessibility across
different population groups. Unlike existing approaches, our methods
are mathematically founded using rigorous optimization and statistical
methodology.

A measurement framework of spatial accessibility must evaluate the
geographic variability of healthcare resources between and within
communities while accounting for constraints in the healthcare delivery
system. Specifically, measures of spatial accessibility need to
consider the number of local physicians from which patients can choose,
the congestion at service sites due to the level and nature of need of
those seeking care, the distance or time impediments to these services,
and the ability of patients to overcome these barriers given their
mobility and socio-economic position [\citet{McGHum09};
Odoki, Kerali and
Santorini (\citeyear{OdoKerSan01})].

% Existing literature on measures for spatial accessibility
The three approaches that have dominated measures of spatial
accessibility are as follows: distance/time to nearest service,
population-to-provider ratios and gravity models [\citet{McGHum09}].
Simple measures such as distance to nearest service site or
population-to-provider ratios are limited in their ability to capture
realistic accessibility patterns because they do not take into account
the trade-off between demand and supply, or patients' decreased
willingness to travel to the further sites [\citet{Kha92}]. To address
this, researchers have attempted to refine the gravity model, which was
inspired by objects' interactions in Newtonian physics and has become
widely used in econometric measurement studies to model spatial
interaction [\citet{TalAns98}]. Recent gravity-based models
account both for individuals' decreasing willingness to travel as
distances increase and for the interaction between distance traveled
and number of people (congestion) at a facility.

% Limitations of the existing measures
All of these measures, including gravity-based models, suffer from one
or more drawbacks. They impose artificial boundaries on individuals'
willingness to travel within catchment service areas, which can lead to
over or underestimates of access depending on the distribution of the
population and regional boundaries. These measures also generally
double-count people and thus overestimate demand in densely populated
areas, or double-count facilities and have the potential to
overestimate supply in areas with a dense network of facilities.
Furthermore, none of these existing methods are capable of measuring
access separately for different populations while still accounting for
the fact that all groups jointly contribute to congestion. Most
importantly, these measures are not well equipped to account for
important aspatial barriers to healthcare, including patients' time
constraints, mobility and ability to pay for services.

% Contributions to measures for spatial accessibility
Mathematically more advanced methods, such as optimization models, are
needed to address these shortcomings and assess the healthcare system's
sensitivity to constraints. By using an optimization model to simulate
the process of patients selecting a physician, we can make assignments
separately for various groups of patients and use constraints to
account for aspatial barriers to healthcare. From the model's output,
we can construct multiple measures to describe various facets of
accessibility, including coverage, travel cost and congestion. Using
the proposed measurement model, we can also evaluate the implications
of changes in the system, like increased physician participation in Medicaid.

These novel aspects in measuring spatial accessibility are essential
for policy evaluation, the ultimate goal of studies of healthcare
access. To further facilitate policy evaluation, we complement the
measurement framework with statistical modeling and inference. First,
we estimate simultaneous confidence bands [Krivobokova, Kneib and
Claeskens (\citeyear{Krivobokova})]
to test for statistical significance of the difference in accessibility
for population groups identified by nongeographic factors, like
insurance status. We then use space-varying coefficient regression
models to estimate the association of various potential explanatory
factors to accessibility [\citet{Ass03}; \citet{Geletal03};
\citet{Waletal07}]. Finally, we assess whether these associations
are space varying and statistically significant with simultaneous
confidence bands [\citet{Ser11}].

When estimating a space-varying coefficient model to make inferences on
spatial accessibility, one challenge is that there are many spatially
varying factors that could potentially be associated with accessibility
and we must consider their effects jointly. Furthermore, these
socioeconomic factors are likely highly collinear. Due to these
difficulties, standard space-varying regression models are instable and
computationally expensive and, therefore, we need to employ advanced
computational methods such as backfitting [Buja, Hastie and
Tibshirani (\citeyear{BujHasTib89})].

% CASE STUDY
Our measurement and modeling framework is applicable to different types
of healthcare specialties and patient populations, and is scalable to
different network densities and varying geographic domains (e.g., state
vs. national). To illustrate the process of implementing this general
framework to study a specific problem, in this paper we focus on
children's accessibility to primary care pediatricians. Primary care
has been acknowledged as the most important form of healthcare for
maintaining population health because it is ``relatively inexpensive,
can be more easily delivered than specialty and inpatient care, and if
properly distributed is most effective in preventing disease
progression on a broad scale'' [\citet{Gua04}; \citet{LuoQi09}].
Pediatric primary care offers health policy makers even more
opportunities, as ``investments during the early years of life have the
greatest potential to reduce health disparities within a generation''
[Marmot et al. (\citeyear{MarFrial08})]. Because childhood poverty is associated with
many health, economic and social problems later in life, we study the
disparities in accessibility between children insured by Medicaid and
other children [\citet{DraRan09}]. We also intend to understand
associations between accessibility and other factors, including income
level, education, unemployment, race, segregation and healthcare
infrastructure. The end goal is to design and evaluate interventions
for increasing spatial accessibility in pediatric healthcare.

%% Structure of the introduction
The remainder of the paper is organized as follows. We present the
optimization model for measuring accessibility along with its
application to policy interventions in Section~\ref{sec:Opt:Model}. In
Section~\ref{sec:Stat:Model} the measurement methodology is
complemented with the analysis of potential contributing factors to
disparities in accessibility using spatial modeling. The proposed
framework for measurement of and inference on accessibility is
investigated in detail for the state of Georgia in Section~\ref
{sec:Georgia}. We summarize our findings and conclusions of this study
in Section~\ref{sec:summary}. Some technical details and additional
simulation studies are deferred to the supplementary material.

%s2 #&#
\section{Modeling framework for measuring accessibility}\label{sec:Opt:Model}

Given the geographic nature of accessibility and availability, any
study of these two dimensions of access requires characterizing the
patient populations and the provider network at the community level. We
represent communities through census tracts, which are designed to
identify homogeneous groups of 2000--8000 people, although some may vary
more widely in population size, particularly in urban areas. The
location of each census tract $i=1,\ldots,S$ is taken to be the
latitude and longitude point of its center of population, which we
denote $s_i$. We use the Environmental Systems Research Institute's
(Esri) ArcGIS software to measure the distance $d_{ij}$ along major
roads between each census tract $i=1,\ldots,S$ and each physician
$j=1,\ldots,T$. The additional characteristics of communities and
healthcare providers needed to study patients' accessibility will
depend on the aspatial barriers to healthcare of the patient groups
under consideration.

We begin our study of accessibility by using a linear optimization
model to describe patients' interactions with the healthcare system.
This model assumes a centralized planner's perspective and assigns
patients to physicians in a manner which maximizes an overall measure
of welfare. While centralized models often fail to account for
important aspects of individuals' behavior, we address this shortcoming
by using constraints to ensure that assignments mimic families' likely
choices given their barriers to healthcare. Because each group of
patients must overcome a unique set of obstacles to obtain healthcare,
different groups are unlikely to make the same decisions about
physicians. To account for these differences and consider their impact
on accessibility, the proposed optimization model makes assignments
separately for each patient group under consideration.

%s2.1 #&#
\subsection{Mathematical modeling of patient behavior}
When considering pediatric healthcare accessibility, the \textit{decision
variables} are $n_{ij}$, or the number of children from census tract
$i$ to assign to physician $j$. In this study, we make assignments
separately for children on Medicaid and for children with other types
of insurance and denote these as $n_{ij}^M$ and $n_{ij}^O$,
respectively. All assignments are nonnegative and limited by each
group's population.

When making these assignments, we assume that families and policy
makers strongly value children having a primary care provider and that,
all else equal, families prefer to visit nearby physicians. We
therefore require that a given percentage of all children are matched
to a provider, and our model's \textit{objective function} minimizes the
total distance patients travel to reach their pediatrician.
Specifically, we make assignments that
\[
\min\sum_{i=1}^S\sum
_{j=1}^T d_{ij} \bigl(n_{ij}^M+n_{ij}^O
\bigr).
\]

We use \textit{constraints} to ensure that the model's assignments
allocate patients among physicians in a realistic manner. For a
physician to remain in practice, he or she must maintain a sufficiently
large number of patients. At the same time, physicians have a maximum
patient capacity ($\mathrm{PC}$) based on the time they must spend with patients
to provide quality care. Assignments must thus satisfy
\[
\mathrm{PC} \times \mathrm{LC} \leq\sum_{i=1}^S
\bigl(n_{ij}^M+n_{ij}^O \bigr) \leq
\mathrm{PC} \qquad\mbox{for any } j=1,\ldots,T,
\]
where $\mathrm{LC}$ represents the lowest level of congestion a physician can
experience and remain in practice. Here, congestion refers to the ratio
$\frac{\mathrm{number\ of\ patients}}{\mathrm{PC}}$. When physicians have higher
congestion, patients have more difficulty\break scheduling appointments,
spend more time in the physician's waiting room, and experience shorter
or rushed visits. Therefore, patients will likely distribute themselves
evenly among their local physicians to avoid high congestion. Some
families will even prefer to drive further away to experience lower
congestion. To capture this behavior, we require that
\[
\sum_{j\in c_i}\sum_{i=1}^S
\bigl(n_{ij}^M+n_{ij}^O \bigr) \leq
\mathrm{PC}\times \mathrm{CC}\times \mathrm{md}_i \qquad\mbox{for any census tract }
c_i \mbox{ with } \mathrm{md}_i\geq2.
\]
Here $\mathrm{md}_i$ is the number of physicians in census track $i$ and $\mathrm{CC}$
represents the maximum congestion likely to occur at the census tract
level in areas with multiple physicians.

We also use constraints to consider \textit{barriers to healthcare}. Here,
we focus on two obstacles especially likely to affect children on
Medicaid: patients' mobility and physicians' willingness to accept new
patients. We assume that there is a maximum distance, $\mathrm{mi}_{\mathrm{max}}$, that
any family is willing or able to travel to reach their primary care
physician. Because families without access to a vehicle either must
walk, use public transportation or rely on others to reach a physician,
they are unlikely to be able to travel as far as other patients. We
therefore enforce the following constraint, which limits families
without cars to physicians within a given small number
($\mathrm{mi}_{\mathrm{max}}^{\mathrm{limited}}$) of miles but allow other families to travel to
any physician within $\mathrm{mi}_{\mathrm{max}}$:
\begin{eqnarray*}
\sum_{j=1}^T n_{ij}^M
I \bigl(d_{ij}\geq \mathrm{mi}_{\mathrm{max}}^{\mathrm{limited}} \bigr) &\leq&
\mathrm{mob}_{i}^M * p_{i}^M,\\
 \sum
_{j=1}^T n_{ij}^O I
\bigl(d_{ij}\geq \mathrm{mi}_{\mathrm{max}}^{\mathrm{limited}} \bigr) &\leq&
\mathrm{mob}_{i}^O * p_{i}^O \qquad\mbox{for any
} i=1,\ldots,S.
\end{eqnarray*}
Here, $p_{i}^M$ and $p_{i}^O$ are the population of children in census
tract $i$ on Medicaid and on other types of insurance, respectively.
$\mathrm{mob}_{i}^M$ and $\mathrm{mob}_{i}^O$ denote the percentage of Medicaid and other
families in census tract $i$ who own at least one vehicle. We consider
Medicaid and other patients' mobility separately to account for the
correlation between income, qualification for Medicaid and car ownership.

Many physicians limit their participation in Medicaid programs due to
the excessive paperwork demands or relatively low reimbursement rates
[Berman et al. (\citeyear{BerDolal02})]. While all patients must consider the
availability of physicians, Medicaid patients therefore must choose
from a smaller pool of physicians. In particular, the model requires
that for each physician,
\[
\sum_{i=1}^S n_{ij}^M
\leq \mathrm{PC}\times \mathrm{MC}_j\times \mathrm{pam}_j \qquad\mbox{for any } j=1,
\ldots,T.
\]
Here, $\mathrm{pam}_j$ is the probability that physician $j$ accepts any
Medicaid patients, and $\mathrm{MC}_j$ is the maximum percentage of physician
$j$'s caseload that he or she is willing to devote to Medicaid patients.

%s2.2 #&#
\subsection{Evaluating policy change}
While there are many policies that could conceivably improve
accessibility, in this paper we consider the effect of interventions
that eliminate or reduce the unique obstacles to healthcare experienced
by Medicaid patients. For example, Medicaid patients' mobility almost
always improves as their rate of car ownership approaches that of other
patients. To consider the effect of policies that improve Medicaid
patients' mobility through public transportation, dial a ride programs
or other means, we evaluate accessibility when
\[
\mathrm{mob}_{i}^{M\prime} = \mathrm{mob}_{i}^M + \lambda
\times\bigl(\mathrm{mob}_{i}^O - \mathrm{mob}_{i}^M
\bigr) \qquad\mbox{for each } \lambda=0,0.05,0.1,\ldots,1.
\]
Here, $\mathrm{mob}_{i}^{M\prime}$ represents Medicaid patients' mobility in
census tract $i$'s after a simulated policy change.

To evaluate the impact of increasing physician participation in
Medicaid programs, we consider two types of policies that we dub as
follows: thresholding and scaling. Thresholding and scaling policies
can affect values of $\mathrm{pam}$ or values of $\mathrm{MC}$. Both types of policies
modify the extent of physician participation in Medicaid programs, but
thresholding policies target areas where Medicaid patients are unserved
while scaling policies affect regions where Medicaid patients are
underserved. While it is difficult to design a policy that would
exclusively have a thresholding or scaling effect, considering the
types of changes separately is useful for analyzing the sources of
limited accessibility for Medicaid populations, and for evaluating the
likely success of interventions. These types of structural changes may
be achieved through policies that increase Medicaid payment levels or
promote Medicaid managed care contracts that allow physicians greater
flexibility [Perloff, Kletke and
Fossett (\citeyear{PerKleFos95})].

We simulate these policies with our optimization model by changing the
values of $\mathrm{pam}_j$ and $\mathrm{MC}$ as follows:

\textit{Thresholding}:
\begin{eqnarray*}
\mathrm{pam}_j^{\prime} &=& \min \{\lambda, \mathrm{pam}_j  \}  \qquad\mbox{for any }
\lambda=0,0.05,0.1,\ldots,1, \\
\mathrm{MC}_j^{\prime} &=& \min \{\lambda, \mathrm{MC}_j  \}\qquad \mbox{for
any }\lambda=0,0.05,0.1,\ldots,1.
\end{eqnarray*}

\textit{Scaling}:
\begin{eqnarray*}
\mathrm{pam}_j^{\prime} &=& \min \{\lambda\times \mathrm{pam}_j, 1 \}
\qquad\mbox{for any } \lambda= 0.5, 0.55, 0.6,\ldots,1,\ldots,2,\\
\mathrm{MC}_j^{\prime} &=& \min \{\lambda\times \mathrm{MC}_j, 1 \}\qquad \mbox{for
any } \lambda= 0.5, 0.55, 0.6,\ldots,1,\ldots,2.
\end{eqnarray*}
Here, $\mathrm{pam}_j^{\prime}$ and $\mathrm{MC}_j^{\prime}$ represent the values of
$\mathrm{pam}_j$ and $\mathrm{MC}_j$ after the simulated policy changes.

%s2.3 #&#
\subsection{Measuring accessibility}
After simulating patient behavior under current and policy-effected
conditions, we use our models' results to construct three measures that
describe important dimensions of children's accessibility. Under each
set of conditions and for each census tract, we evaluate the following:
\begin{longlist}
\item[\textit{Coverage}] describing children's ability to find physicians
who will serve them and derived as follows,
$
C_i = \frac{1}{p_i} \sum_{j=1}^T  (n_{ij}^M+n_{ij}^O )$;
\item[\textit{Travel Cost}] defining the average distance children must
travel to reach their physician and derived as follows, $
\mathit{TC}_i = \frac{1}{p_i} \sum_{j=1}^T d_{ij}
(n_{ij}^M+n_{ij}^O )
+ \mathrm{mi}_{\mathrm{max}}*(1-C_i)$;
\item[\textit{Congestion}] measuring the average congestion that patients
experience for their physician and derived as follows,
$
\mathit{CG}_i = \frac{1}{p_i}  [\sum_{j=1}^T
(n_{ij}^M+n_{ij}^O
)\times\break \frac{\sum_{i=1}^S (n_{ij}^M+n_{ij}^O )}{\mathrm{PC}} ] + (1-C_i)$.
\end{longlist}
Here $p_i$ is the total child population in census tract $i$, so $p_i =
p_{i}^O + p_{i}^M$. Note that because children who are not served by a
pediatrician have the worst possible accessibility, regions where $C_i
= 0$ are assumed to experience a travel cost of $\mathrm{mi}_{\mathrm{max}}$ miles and
100\% congestion. We adjust these formulas to separately evaluate
Medicaid and non-Medicaid patient's accessibility.

Evaluating these various dimensions can help policy makers determine a
strategy that best suits their objectives. To estimate a policy's
impact, we consider how each group's population-based accessibility
changes as the policy is gradually implemented. The changes are
evaluated for each of the three dimensions (coverage, travel cost and
congestion) and at different aggregation levels (census tract
variations vs. state wide). A given policy can improve accessibility
for the overall population or it can improve accessibility for one
group at the expense of another. Policies may reduce congestion and
increase travel times or vice versa. Some policies will have a small
impact in many areas, while others will have a more substantial effect
on a smaller number of regions. In this research, we target policies
that are (approximately) Pareto optimal in the sense that they improve
some dimensions of accessibility for some groups without significantly
reducing the accessibility of other populations [\citet{MarSch94}].

%s3 #&#
\section{The equity of healthcare accessibility}\label{sec:Stat:Model}
Our second primary research focus is developing a framework for
understanding the equity of healthcare accessibility as it relates to
geographic, demographic, socioeconomic and healthcare infrastructure
variables. Following \citet{BraGru03}, we define \textit{equity}
as the absence of systematic disparities in accessibility
between different groups of people, distinguished by different levels
of social advantage/disadvantage. More specifically, equity is achieved
when the expectation of accessibility given potential contributing
factors to inequities is (approximately) equal to the expected
accessibility in the population unconditional of any contributing
factor [\citet{FleSch09}]. Mathematically speaking, if
$Y$ is the spatial accessibility and $X$ is the set of observed
contributing factors, equity is achieved when the expectation of the
conditional distribution $Y|X$ is equal to the expectation of the
marginal distribution of $Y$. Practically speaking, in an equitable
system, no systematic association will be found between spatial
accessibility and the independent factors. We use statistical models to
estimate the associative relationships between the dependent variable,
accessibility and the potential contributing factors, and to test if
these associations are statistically significant. Importantly,
statistically significant and spatially nonrandom association patterns
between a particular factor and the accessibility metric indicate
potential inequities with respect to that variable.

%s3.1 #&#
\subsection{Determining the effect of participation in Medicaid}

We begin our study of inequities in accessibility by using statistical
hypothesis testing to consider the association between spatial
accessibility and participation in Medicaid programs. Here, we denote
the $i$th census tract's accessibility measures for the Medicaid and
other population as $M(s_i)$ and $O(s_i)$, respectively. Because
$M(\cdot)$ and $O(\cdot)$ are spatial processes that are measured for
the same set of units, we can take their difference $Z(s_i) =
M(s_i)-O(s_i)$ for $i=1,\ldots,S$. If there is not a significant
difference between these populations' accessibilities, $Z(s)$ is
approximately zero regardless of the spatial location. We translate
this as a hypothesis testing problem where the null hypothesis is
$H_0\dvtx Z=0$ across the geographic domain and the alternative
hypothesis is
$H_1\dvtx Z(s)\neq0$ for some areas within the geographic domain. We use
nonparametric methods to derive our decision rule for this test. If we
reject the null hypothesis, we conclude that there are regions where
accessibility for Medicaid patients differs from that of other
patients. Based on this procedure, we can also create a \textit{significance map}
that identifies specific locations where the
difference in the two populations' accessibility is statistically
different from zero. We provide details on how to proceed with this
inference method in Supplementary Material A.

%s3.2 #&#
\subsection{Determining the effect of geographic location}\label{sec3.2}

This type of statistical hypothesis testing can also be used to
consider the association between patients' accessibility and the
location of their homes. To identify locations where accessibility is
statistically significantly different than that of the overall region,
we test the null hypothesis that $H_0\dvtx Y = {\mu_{0Y}}$ across the
geographic domain vs. the alternative hypothesis that $H_1\dvtx Y(s)
\neq
\mu_{0Y}$ for some areas within the geographic domain. Here, $Y(s_i)$
is a measure of spatial accessibility for census tract $i$, and $\mu
_{0Y}$ is some equity threshold. For example, in this paper $\mu_{0Y}$
is the population-weighted average of $Y(s_i)$ across locations
$i=1,\ldots, S$. While the resulting significance map identifies the
most underserved locations, further interpretation of these results is
challenging because setting $\mu_{0Y}$ is quite subjective. We
therefore turn our focus to statistical methods that can more precisely
characterize the association between a wide set of geographic and
socio-economic factors and spatial accessibility.

%s3.3 #&#
\subsection{Contributing potential factors to spatial accessibility}
\label{subsection:Factors}

While there are an unlimited number of factors that could potentially
affect accessibility, we focus on factors that have been previously
linked to limited accessibility, especially for vulnerable populations,
like Medicaid patients.

Economic and racial factors are commonly cited in the literature as
predicting physician participation in Medicaid, which impacts all
groups' accessibility [Hambidge et al. (\citeyear{HamEmsal07}); \citet{WanLuo05}]. We
use three factors to consider the economic climate of the census tract:
the \textit{median household income}, the \textit{unemployment rate}
and the \textit{percentage of adults who have an associate, bachelor or
graduate degree}. We use the percent of the population that is nonwhite
to evaluate the \textit{racial composition} of the tract. Some papers
argue that the amount of \textit{segregation} in a community has more
impact on physician participation in Medicaid than race [\citet{FisWil04}; \citet{WilMoh09}]. We therefore also
consider a segregation measure that compares the diversity in local
neighborhoods to diversity in broader communities as suggested by
\citet{Reaetal08}. Further details on this segregation measure are
given in Supplementary Material A.

The structure of the provider network may also affect patients'
accessibility. Because an area's \textit{distance to hospitals} and its
\textit{population density} affect its market size, these factors may
influence where physicians choose to practice. To measure census tract
$i$'s distance to hospitals, we take the average distance from $s_{i}$
to all hospitals within 25 miles, weighted by the size of the hospital
as measured by its number of beds. Traditional measures of population
density are estimated by dividing the population of a census tract by
its land area. This type of density measure does not account for the
irregular shapes and sizes of census tracts, and ignores the spatial
dispersion of the population from a census tract to its neighbors. A
more appropriate method is to assume that the population forms a point
process with a spatially varying rate and estimate its density using
nonparametric density estimation. As we describe in Supplementary
Material A [\citet{supp}], we use the classical
Kernel Density Estimation (KDE) method, which is one of the most widely
used methods for this purpose [\citet{Dig85}] and is known to be a
consistent estimator [\citet{Par62}].

In our subsequent statistical models, the independent variables will be
chosen from the seven factors described in this section.

%s3.4 #&#
\subsection{The space-varying coefficient model}

One difficulty in estimating the association between accessibility and
the potential explanatory factors is that these factors may vary over
the geographic space systematically, that is, they may display
nonrandom geographic patterns. Furthermore, the unknown relationship
between accessibility and the explanatory factors may also vary with
the geographic space in a nonrandom fashion. This suggests spatially
varying coefficients in a regression setting. Varying coefficient
regression models have been applied to longitudinal data to estimate
time-dependent effects of a response variable [\citet{FanZha00};
\citet{HasTib93}; \citet{Hooetal98}; Wu and Liang
(\citeyear{WuLia04})] and to spatial data [\citet{Ass03}; \citet{Geletal03};
\citet{Waletal07}]. Because our spatial domain is densely sampled,
we can apply this model to estimate spatial association maps between
accessibility and the explanatory variables.

A second difficulty, as highlighted in the previous section, is that
models which evaluate sources of inequities will include a large number
of explanatory variables. To address this challenge, we employ an
estimation algorithm that uses partial residual fitting or backfitting
[Buja, Hastie and
Tibshirani (\citeyear{BujHasTib89})]. Furthermore, spatial collinearity among the
explanatory variables leads us to view each model as an approximation
to an unknown, true model and to conjecture that multiple models can
capture the associative relationships to the accessibility measure. We
therefore do not focus on selecting a single, best model, and instead
develop a procedure for systematically evaluating multiple models,
where each model includes a different set of factors that can each take
a constant or nonconstant shape. We seek consistency in the statistical
significance and the shape of the regression coefficients across these
models to capture the essential relationships between the factors and
accessibility.

%s3.5 #&#
\subsection{Estimation}
Space-varying coefficient models assume that the response variable
$Y_{i} = Y(s_i), i=1,\ldots, S$ observed at location $s_i$ is explained
by a set of covariates $ (X_{r,i} = X_r(s_i); r=1,\ldots,R )$
such that
\[
\mathbb{E}[Y_{i}|X] =\beta_{1}(s_i)X_{1,i}+
\cdots+ \beta_{R}(s_i)X_{R,i},
\]
where $\beta_{r}(s)$ for $r=1,\ldots,R$ are smooth coefficient
functions over a geographic space $s\in{\cal S}$. For example, in our
studies, the locations $\{s_i\}$ correspond to the population centers
of the 1618 census tracts in Georgia and $\{X_{r,i}; r=1,\ldots,7\}$ is
the set of geographic and socioeconomic factors described in
Section~\ref{subsection:Factors}.

Since the regression coefficients $\beta_{r}(s)$ for $r=1,\ldots,R$ are
unknown functions, we use nonparametric methods to estimate them.
Specifically, we decompose
%
%e1 #&#
\begin{equation}
\label{eq:decom} \beta_r(s) = \sum_{k=1}^{K_r}
\theta_{rk} \phi_k(s),
\end{equation}
where $\{\phi_1(s),\phi_2(s),\ldots\}$ is an orthogonal basis of
functions in $L^2({\cal S})$ and $\theta_{rk}, k=1,\ldots, K_r$ are
unknown parameters. The number of basis functions used in the
decomposition, $K_r$, controls the smoothness of the function $\beta
_r(s)$. If $K_r$ is small, the estimated function is very smooth,
resulting in a larger estimation bias, whereas if it is large, the
estimated function is highly variable, resulting in overfitting.
Therefore, the selection of $K_r$ is important; if we do not use an
optimal value, the estimated association patterns may reveal spurious
associations.

To address the challenge of selecting the $K_r$'s without increasing the
computational effort, we estimate the space-varying coefficients using
penalized splines [Ruppert, Wand and
Carroll (\citeyear{RupWanCar03})]. In penalized spline
regression, $K_r$ is chosen to be sufficiently large to ensure a small
modeling bias [\citet{LiRup08}], but constraints are imposed on
the coefficients $\theta_{rk}, k=1,\ldots,K_r$ through a penalty
function $J(\beta_r(s)) = J (\{\theta_{rk}\}_ {k=1,\ldots
,K_r}
)$ to limit the influence of $\theta_{rk}, k=1,\ldots,K_r$ and control
the smoothness of the regression coefficients. As further described in
Supplementary Material B, this is equivalent to estimating the
coefficients using penalized regression [\citet{supp}].

Because of difficulties in evaluating space-varying coefficient models
with many predictors, we also borrow an estimation idea that has been
previously used in the generalized additive model and other fitting
algorithms [Buja, Hastie and
Tibshirani (\citeyear{BujHasTib89}); \citet{HasTib90}]. These
algorithms rely on partial residual fitting and are conceptually
similar to the Newton Raphson algorithm that successfully fits a large
number of nonlinear equations by iteratively solving one equation or
parameter at a time until the solution converges. We mimic this
procedure and estimate the association coefficients using the
Backfitting algorithm described in Supplementary Material B [\citet{supp}].

%s3.6 #&#
\subsection{Inference and policy implications}

To interpret our results, we first assess the significance and shape of
the estimated association coefficients. Coefficients have two possible
shapes: constant and nonconstant. Factors with constant shaped
coefficients influence accessibility in the same manner in all
locations, while factors with nonconstant shaped coefficients have an
association with accessibility that varies across regions.

In regression models, hypothesis testing is the common procedure for
assessing the significance and shape of the coefficients. Specifically,
we are interested in the results of the following hypothesis test for
each of the $r$ coefficients:
\[
H_{r0}\dvtx \beta_r(s) = c\quad \mbox{vs.} \quad H_{r1}
\dvtx \beta_r(s)\qquad \mbox{nonconstant},
\]
where $c$ is a constant value. If the null hypothesis is not rejected,
it is plausible that the corresponding coefficient is constant and
further tests should be conducted to determine if this coefficient is
statistically significant, that is, $c\neq0$.

Inspired by \citet{Ser11}, we propose identifying the shape of the
coefficients using confidence bands rather than hypothesis testing.
Specifically, if $\mathrm{CB}_{\alpha}$ is a $1-\alpha$ simultaneous confidence
band for the coefficient $\beta_r(s)$, then $P(\beta_r(s)\in
\mathrm{CB}_{\alpha
}, s\in{\cal S})\geq1-\alpha$ where ${\cal S}$ is the space domain.
We decide that the coefficient is not constant if there does not exist
a constant plane $p(s)=c,\ \forall s$ such that $p(s)\in \mathrm{CB}_{\alpha}$.

Based on the confidence bands, we also examine the statistical
significance of the coefficients and construct positive and negative
\textit{significance maps}. A positive (negative) significance map
consists of spatial regions that have a statistically significant
positive (negative) association between access and the corresponding
explanatory variable. \textit{The presence of broad regions of
positive or
negative significance in such a map is an indication of potential
inequities for the corresponding explanatory variable.}

The shape and significance of socioeconomic or geographic factors'
association coefficients should be considered when policy makers design
interventions to combat inequities. If a factor has a nonconstant
effect on accessibility, an intervention that reduces accessibility in
some areas may increase accessibility in others, and so the instrument
that policy makers use should vary across locations. Furthermore, while
all variables that have a nonconstant effect on accessibility have a
significant impact in at least one location, there are often many areas
where there is no significant relationship between the nonconstant
factor and accessibility. While devising policies around constant
factors is simpler, this is also not without challenges. For example,
these policies should not necessarily be applied uniformly across the
state because regions with large populations of the given demographic
may offer policy makers more ``bang for their buck.''

%s3.7 #&#
\subsection{Model evaluation}

We evaluate the full model and a large number of reduced models, which
each include four or more variables. Models that perform well meet
three criteria: (1) a small AIC value, (2) a small correlation between
residuals and the accessibility measure and (3) a small Moran I
statistic value on model residuals. The second criterion is used as a
measure of how well we explain the accessibility measure with the
explanatory factors included in the model. For example, \citet{TibKni99}
relate this measure to the coefficient of determination.
We estimate the spatial correlation between two processes similarly to
Jiang (\citeyear{Jian10}). The third criterion is a measure of how much spatial
dependence is left in the residuals. From the large set of models
considered, we selected only those that simultaneously show an
improvement over all three criteria.

As mentioned earlier, because our factors are spatially collinear, we
do not attempt to identify a single, best model, but instead analyze
the consistency of many models' results. To describe the relationship
of a specific predictor with the accessibility measure, we focus on
three characteristics of the corresponding association coefficient: its
shape, its significance and the range of its values. If these
properties are consistent across many models, we conclude that our
models have captured the underlying relationship between this factor
and accessibility. Consistency across models also suggests that the
full set of factors collectively explains patterns of accessibility.
Conversely, wide variations across the models' results may indicate
that there are important determinants of access that are not included
in the model.

%s4 #&#
\section{Pediatric accessibility in Georgia}\label{sec:Georgia}

We pilot the previous sections' methodology on pediatric primary care
for Georgia, one of the 10 worst states for many measures of child
health [Kids Count National Indicators (\citeyear{KC10})]. To implement our models, we
need a broad and detailed set of data to describe the characteristics
of patients and physicians in Georgia. Demographic information about
Georgia's population was acquired from many sources, including the
Census Bureau. The addresses of the primary care pediatricians located
in Georgia were acquired from the Centers of Medicare and Medicaid
Services' (CMS') National Provider Identifier (NPI) Registry.

We also use data to select appropriate values for the parameters in the
accessibility measurement model. Because we assume that all patients
are entitled to the same level of care, some of these parameters should
be constant across all patients or all physicians. For example, the
U.S. Department of Health and Human Services defines Medically
Underserved Areas (MUA) as regions with no primary care providers
within 25 miles. To follow these guidelines, we assume that $\mathrm{mi}_{\mathrm{max}}$,
or the maximum distance any patient is willing to travel, is 25 miles.
While we believe that families living in rural areas may currently
travel close to this distance or even slightly further to reach a
pediatrician, we do not consider different maximum travel distances for
various populations. Doing so would result in inequitable estimates and
distort our conclusions. Similarly, we assume that
$\mathrm{mi}_{\mathrm{max}}^{\mathrm{limited}}$, or the maximum distance patients without cars are
willing to travel, is the same for all populations. Other parameters
that should remain constant are $\mathrm{PC}$, physicians' maximum patient
capacity, $\mathrm{LC}$, the lowest level of congestion a physician can
experience and remain in business, and $\mathrm{CC}$, the maximum congestion
level likely to occur in census tracts with multiple physicians.

While these parameters reflect the ease and quality of patient's access
to care, other parameters describe the more basic structure of the
existing healthcare provider network. Two such parameters are $\mathrm{pam}_{j}$
and $\mathrm{MC}_{j}$. These parameters describe the probability that physician
$j$ accepts any Medicaid patients and the maximum percent of physician
$j$'s caseload that he or she is willing to devote to Medicaid
patients, respectively. These parameters will vary across physicians
and our model should reflect this in order to capture accurate
information about the medical services available to children enrolled
in Medicaid programs. Table~\ref{table:parameters} gives further detail
on the values of the parameters in our model, and a more in depth
description of our data sources is provided in Supplementary Material C
[\citet{supp}].

%t1 #&#
\begin{table}
\caption{The variability, values and data sources for the seven
parameters used in our model of accessibility}\label{table:parameters}
\begin{tabular*}{\textwidth}{@{\extracolsep{\fill}}lccc@{}}
\hline
\textbf{Parameter} & \textbf{Variable} & \textbf{Value(s)} & \textbf{Data source} \\
\hline
$\max_{\mathrm{mi}}$ & No & 25 & U.S. DHHS \\
$\max_{\mathrm{mi}}^{\mathrm{limited}}$ & No & 10 & -- \\
$\mathrm{PC}$ & No & 2500 & U.S. DHHS \\
$\mathrm{LC}$ & No & 0.25 & -- \\
$\mathrm{CC}$ & No & 0.70 & -- \\
$\mathrm{pam}_{j}$ & Yes, & $[0,1]$ & Georgia Board \\
& by county & & of Physicians \\
$\mathrm{MC}$ & Yes, & 0.74 if in public hospital & American Academy \\
& by practice setting & 0.64 if in community health clinic & of
Pediatrics \\
& & 0.32 in other setting & \\
\hline
\end{tabular*}
\end{table}

%s4.1 #&#
\subsection{Implementing the accessibility measurement model given
limited data}
Table~\ref{table:parameters} highlights the fact that there are limited
data available to describe many aspects of our accessibility model. To
compensate for these shortcomings, we perform sensitivity analysis to
determine if the model's results are heavily dependent on the value of
each parameter. Figures~2--5 in the supplementary material show that the
population-weighted average, state-wide accessibility, is not highly
sensitive to subtle changes in $\mathrm{PC}$, $\mathrm{LC}$, $\mathrm{CC}$ and
$\mathrm{mi}_{\mathrm{max}}^{\mathrm{limited}}$ [\citet{supp}]. Furthermore, as
these parameters vary across reasonable levels, Medicaid and other
patients' accessibility change at similar rates. More details on these
results are given in Supplementary Material D [\citet{supp}]. Given these conclusions, we are comfortable proceeding with
our assumed values for these parameters.

A second tool for addressing data limitations is a simulation study.
This is an especially useful method when a parameter is expected to
vary across populations and only the parameter's (estimated)
distribution over the population is available. In this case, we sample
from the expected distribution to simulate the parameter value for each
population or region. We then repeat this process multiple times and
consider the mean and variance of the resulting accessibility measures.
In some sense, this simulation algorithm uses the idea of Monte Carlo
uncertainty analysis in Bayesian estimation, where the estimated
distribution plays the role of the prior distribution entering in the
accessibility measurement model.

An example of the application of the simulation idea is in the
specification of $\mathrm{pam}_{j}$ parameter values. Realistically, this
parameter will vary by physician and only takes two possible values:
one if the physician participates in Medicaid programs, and zero if he
does not participate in Medicaid programs. This level of detail is not
present in publicly available data, which only specify the county-level
percentage of physicians accepting any Medicaid patients. We use these
percentages to specify the Bernoulli distribution from which we draw
samples for $\mathrm{pam}_{j}$. We repeat this process twenty times and find
that the variability in the resulting accessibility measures across
trials is very small in most regions. This implies that subtle changes
in the network of physicians participating in Medicaid programs will
likely not have a significant impact on accessibility in most regions.
Supplementary Material D presents figures and further discussion of
these results [\citet{supp}]. In our regression
models, the dependent variables are the means of these simulated
measures of accessibility.

%s4.2 #&#
\subsection{Evaluating the current state of accessibility}

\begin{figure}
\centering
\begin{tabular}{@{}cc@{}}

\includegraphics{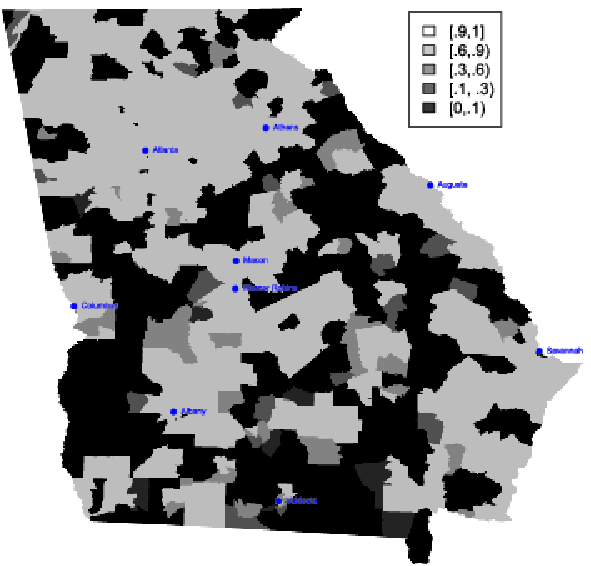}
 & \includegraphics{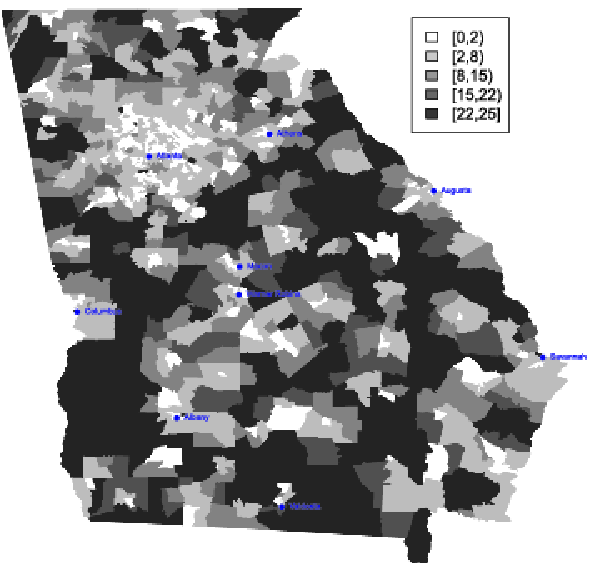}\\
\footnotesize{(a) Coverage} & \footnotesize{(b) Travel cost}
\end{tabular}\vspace*{3pt}
\centering
\begin{tabular}{@{}c@{}}

\includegraphics{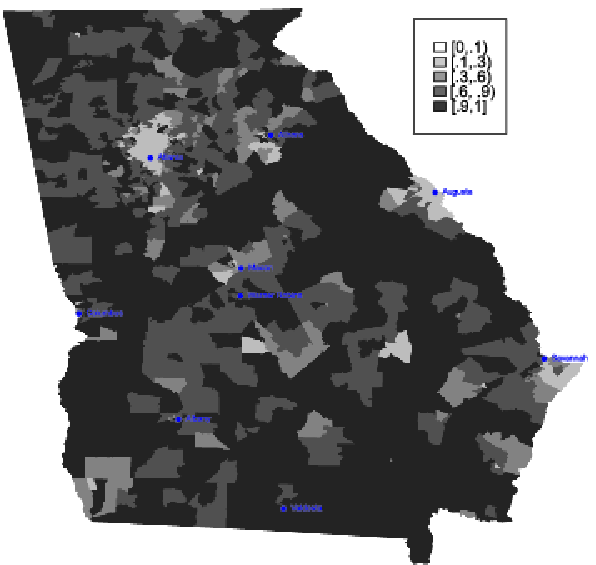}
\\
\footnotesize{(c) Congestion}
\end{tabular}
\caption{Estimated accessibility of children to primary care
pediatricians, as measured by coverage rates, travel costs and
congestion. Each measure of accessibility is derived using the proposed
optimization approach. The maps show estimated accessibility in 2010 in
census tracts in the state of Georgia.}
\label{fig::base}
\end{figure}

Figure~\ref{fig::base} shows the current state of accessibility for the
overall population of children in Georgia. Coverage is nearly 100\% in
broad regions surrounding the most populated cities and towns, but is
nearly 0\% in many rural areas, especially in the southern portion of
the state. We find that travel cost tends to be very high in these
regions where coverage is low. On the other hand, in areas with high
coverage, the distance families must travel to reach their pediatrician
is often below 5 miles and rarely above 15 miles. Together, these
observations suggest \textit{a dichotomy in these two dimensions of
accessibility: either a family is served by a pediatrician located
close to their home or they struggle to find any pediatrician that
meets the minimum standards for accessibility.}

%f1 #&#

With the exception of Atlanta, the state's largest metro area, and
Augusta, the location of the state's medical school, there are only a
handful of regions in the state with excellent accessibility across all
three measures. Many of the rural and suburban areas that benefit from
high coverage and low travel distances encounter high congestion, which
highlights the importance of considering multiple dimensions of
accessibility. High congestion can occur in rural areas if the region
has a low supply of physicians relative to its population, or if the
region is classified as a medically underserved area (MUA) and assumed
to have the worst possible accessibility.

%s4.3 #&#
\subsection{Evaluating policy interventions} \label{subsection:GAPolicy}
\begin{figure}[b]
\centering
\begin{tabular}{@{}cc@{}}

\includegraphics{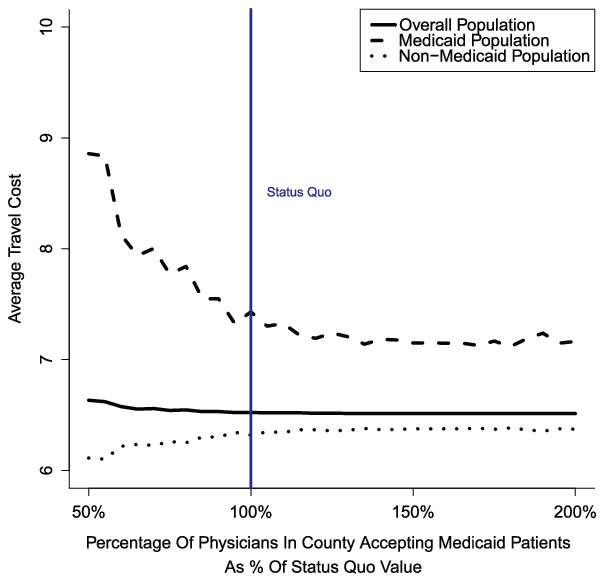}
 & \includegraphics{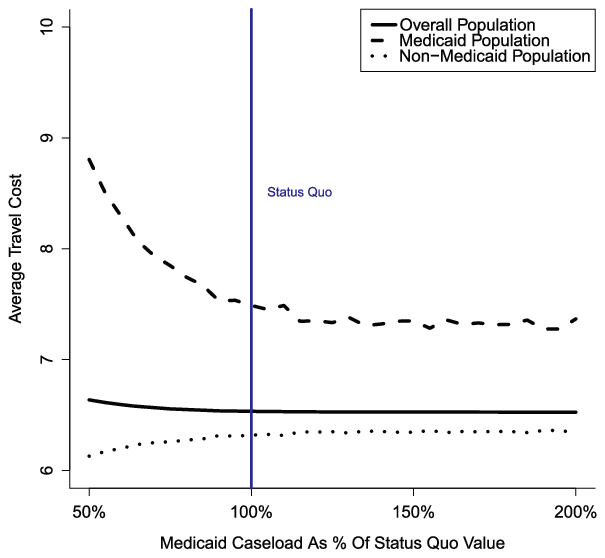}\\
\footnotesize{(a) Effect of changing {pam}} & \footnotesize{(b) Effect of changing {MC}}
\end{tabular}
\caption{The impact of scaling policies on children's state-wide,
population-weighted average distance traveled to reach primary care
pediatricians in Georgia. \textup{(a)} shows the effects of policies that
change the percentage of physicians accepting any Medicaid patients,
while \textup{(b)} shows the effects of policies that change the proportion
of physicians' caseloads that they are willing to devote to Medicaid patients.}
\label{fig::scaling_policy_distance}
\end{figure}

Of the potential changes to the healthcare system that we consider,
\textit{reductions in physicians' participation in Medicaid programs have
the most significant effect on patients' accessibility.} Table~\ref
{table:reduction_policy_changes} and Figure~\ref
{fig::scaling_policy_distance} show that policies that reduce the
number of physicians accepting Medicaid patients and policies that
limit pediatricians' Medicaid caseloads have similar impacts on
Medicaid patients' average accessibility. Both policies cause Medicaid
patients' average travel cost to increase by nearly 20\%. Reducing
$\mathrm{pam}$ has a slightly positive effect on Medicaid patients' congestion,
and reducing $\mathrm{MC}$ has a more substantial positive effect. One
interpretation of these seemingly counterintuitive finding is that
these policies reduce physicians' total caseloads by restricting
Medicaid patients' access. Indeed, Medicaid patients' coverage rates
decrease by approximately 6 percentage points under these policies.
This result also suggests that the burden of these policies may be
unequally distributed among Medicaid patients, with some Medicaid
patients experiencing little change in their accessibility and others
finding that they now have no options for accessible medical care.

%t2 #&#
\begin{table}
\tabcolsep=0pt
\caption{This table shows the estimated effect of policies that reduce
physicians' participation in Medicaid programs. These polices and the
interpretation of $\lambda$ are described in more detail in
Section~\protect\ref{sec3.2}. For each policy, the first row shows the change in Georgia's
Medicaid patients' state-wide, population-weighted average coverage
rate, travel cost and congestion as one moves from the status quo to the
maximum simulated reduction. The second row gives the percent change of
these values}
\label{table:reduction_policy_changes}
\begin{tabular*}{\textwidth}{@{\extracolsep{\fill}}lcccc@{}}
\hline
 &  & \textbf{Coverage rate} & \textbf{Travel cost} & \textbf{Congestion} \\
\textbf{Policy}& $\bolds{\lambda}$& \textbf{of Medicaid patients} & \textbf{of Medicaid patients} & \textbf{of Medicaid patients}
\\
\hline
$\mathrm{MC}$ scaling & $1 \rightarrow0.5$ & $85.0\% \rightarrow78.8\%$ & $7.5
\rightarrow8.8$ & $66.9\% \rightarrow64.7\%$ \\
& & $(-7.2\%)$ & (17.6\%) & $(-3.2\%)$ \\
$\mathrm{pam}$ scaling & $1 \rightarrow0.5$ & $85.3\% \rightarrow79.6\%$ &
$7.4 \rightarrow8.9$ & $67.2\% \rightarrow66.7\%$ \\
& & $(-6.7\%)$ & (19.2\%) & $(-0.6\%)$ \\
\hline
\end{tabular*}
\end{table}

%f2 #&#

Table~1 and Figures~6--8 in Supplementary Material D show that none of
the initiatives designed to improve Medicaid patients' accessibility
deliver on their promise [\citet{supp}]. The
inability of these policies to create significant change in access to
healthcare indicates that spatial accessibility may be limited by the
current distribution of pediatricians. Among the 159 counties in
Georgia, approximately $1/3$ have no pediatrician. Given this network of
pediatricians, unless policies succeed in sending medical professionals
to underserved areas, they are unlikely to accomplish substantial
change. Examples of such policies include programs which forgive
physicians' graduate medical education loans in exchange for their
practice in rural areas for a given period of time
(\surl{http://gbpw.georgia.gov/loan-repayment-programs}). Physicians may also
be able to serve patients with limited accessibility through
telemedicine initiatives [Marcin et al. (\citeyear{MarEllal04})]. Measuring the effects
of these types of policies on accessibility is an interesting direction
for future research.

%s4.4 #&#
\subsection{Determining the effect of Medicaid participation on
accessibility in Georgia}

\begin{figure}
\centering
\begin{tabular}{@{}cc@{}}

\includegraphics{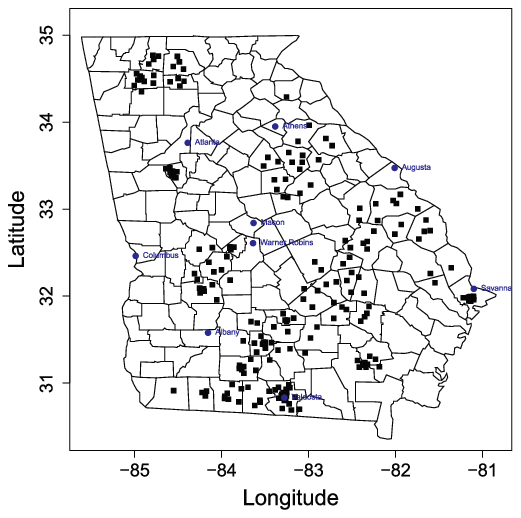}
 & \includegraphics{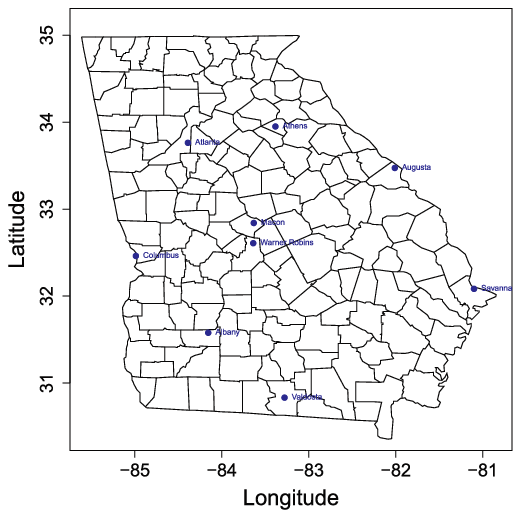}\\
\footnotesize{(a) Locations where other population} & \footnotesize{(b) Locations where Medicaid population}\\[-2pt]
\footnotesize{has significantly higher coverage} & \footnotesize{has significantly higher coverage}\\[3pt]

\includegraphics{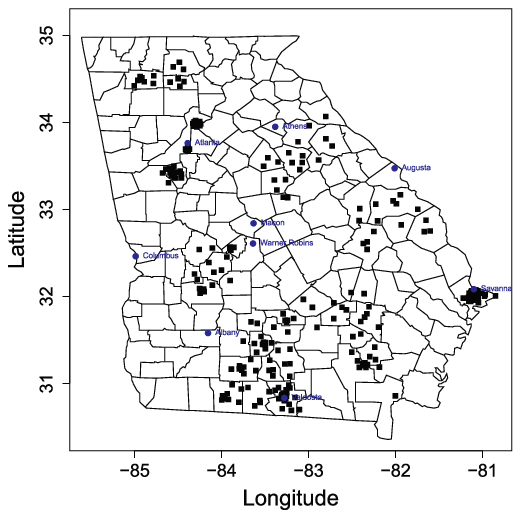}
 & \includegraphics{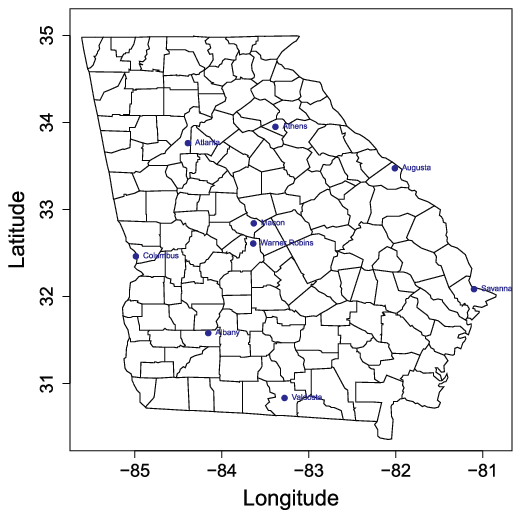}\\
\footnotesize{(c) Locations where other population} & \footnotesize{(d) Locations where Medicaid population}\\[-2pt]
\footnotesize{has significantly lower travel cost} & \footnotesize{has significantly lower travel cost}\\[3pt]

\includegraphics{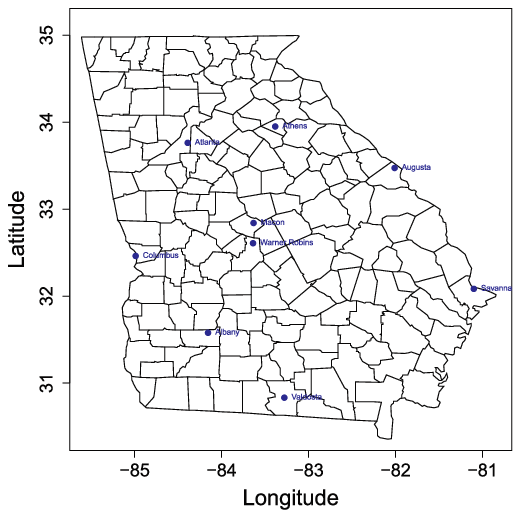}
 & \includegraphics{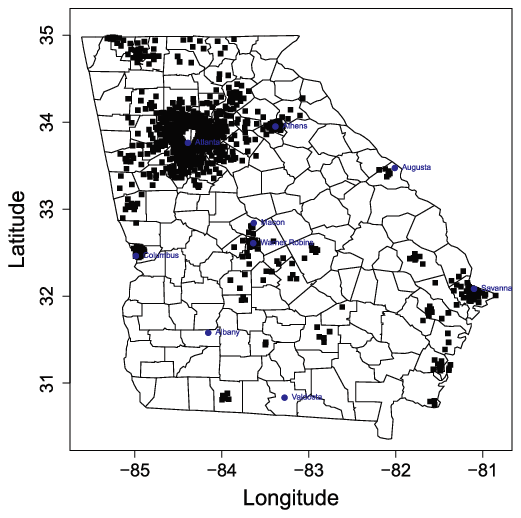}\\
\footnotesize{(e) Locations where other population} & \footnotesize{(f) Locations where Medicaid population}\\[-2pt]
\footnotesize{has significantly lower congestion} & \footnotesize{has significantly lower congestion}
\end{tabular}
\caption{The points on these maps are the centroids of census tracts
where there is a statistically significant difference in the
accessibility of Medicaid patients and other patients.}
\label{fig::Effect_Of_Medicaid}
\end{figure}

Figure~\ref{fig::Effect_Of_Medicaid} shows locations where one
population has statistically significantly better accessibility than
the other. As we would expect a priori, in many rural census tracts,
other patients have significantly higher coverage and lower travel
costs than Medicaid patients. Furthermore, there are no areas where
Medicaid patients experience these advantages. We find that the
direction of advantage is reversed for congestion, and there are many
urban census tracts where Medicaid patients experience significantly
lower congestion than other patients. Many physicians prefer privately
insured patients to Medicaid patients and may be less likely to accept
Medicaid patients if they can afford to only serve patients with
private insurance. In regions where there is a high supply of
physicians relative to the population, congestion is likely to be low
and physicians may have greater incentive to accept all types of
patients [\citet{PerKleFos95}]. Therefore, if Medicaid
patients are able to be served by a physician, they may be more likely
to experience lower congestion. Our results suggest that this theory
holds true in Georgia.

%f3 #&#
%% \figlink{}
%

%
%Coverage]
%{\includegraphics[width=60mm]{plots/Other_Advantage_Coverage.pdf}}
%Coverage]
%{\includegraphics[width=60mm]{plots/Medicaid_Advantage_Coverage.pdf}}
%
%Travel Cost]
%{\includegraphics[width=60mm]{plots/Other_Advantage_Distance.pdf}}
%Travel Cost]
%{\includegraphics[width=60mm]{plots/Medicaid_Advantage_Distance.pdf}}
%
%Congestion]
%{\includegraphics
%[width=60mm]{plots/Other_Advantage_Congestion.pdf}}
%Congestion]
%{\includegraphics[width=60mm]{plots/Medicaid_Advantage_Congestion.pdf}}
%

Figure~9 in Supplementary Material D shows how these maps would change if
physicians were to reduce their Medicaid caseload ($\mathrm{MC}$) by fifty
percent [\citet{supp}]. Under current conditions,
there are 263 census tracts in Georgia where Medicaid patients
experience statistically significantly higher travel costs than other
patients. If policy changes prompt physicians to make these reductions
in their Medicaid caseloads, this number would increase to 360. While
this increase is considerable, there are also many communities whose
access is unaffected by this shift in physicians' attitudes. These
findings further support our analysis in Section~\ref{subsection:GAPolicy}.

%s4.5 #&#
\subsection{Factors associated with spatial accessibility in Georgia}
Figure~10 in Supplementary Material D shows that accessibility is
likely to be significantly higher than the population-weighted
state-wide average in urban areas of Georgia, and significantly lower
in rural regions of Georgia [\citet{supp}]. This is
especially true for coverage and travel cost. To determine if there are
factors associated with these geographical disparities in
accessibility, we apply the regression approach described in
Section~\ref{sec:Stat:Model}. Here, we run two sets of models: one
where the
dependent variable is the travel cost of the overall population and one
where the dependent variable is the travel cost of the Medicaid
population. The independent variables are chosen from the set of seven
factors described in Section~\ref{subsection:Factors}. Table~\ref
{table:models} summarizes the results for the overall population based
on the estimation of multiple models, each with four or more
independent variables.

%t3 #&#
\begin{table}
\caption{The consistency, shape (constant vs. nonconstant), statistical
significance and coefficient values observed across multiple models of
the overall population's travel cost in Georgia. These models each
contain different combinations of the seven explanatory factors
considered in the association analysis}
\label{table:models}
\begin{tabular*}{\textwidth}{@{\extracolsep{\fill}}lcccc@{}}
\hline
\textbf{Factor} & \textbf{Consistent} & \textbf{Significant} &
\textbf{Shape} & \textbf{Range} \\
\hline
Median household & Yes & Yes & Constant & [0.09, 0.37]\\
Income & & & & \\
Percent with higher education& No & Yes & Constant $\&$ & $[-0.29, -0.26]$\\
 & & & nonconstant & N/A \\
Unemployment rate & Yes & No & Constant & [0.026, 0.21]\\
Percent of nonwhite population& Yes & No & Constant & $[-0.18, 0.63]$\\
% & & & & \\
Population density & Yes & Yes & Nonconstant & N/A \\
Distance to hospitals & No & Yes & Constant $\&$ & [0.14, 0.16] \\
& & & nonconstant & N/A \\
Diversity ratio & Yes & Yes & Constant \& & $[-0.23, -0.19]$ \\
& & & nonconstant & N/A \\
\hline
\end{tabular*}
\end{table}

While the results in Table~\ref{table:models} confirm that most factors
commonly cited in the literature do indeed have a significant
relationship with both populations' spatial accessibility, there are
two noteworthy exceptions: unemployment and race. We find more support
for the emerging line of research which argues that the amount of
racial diversity in a community is likely to be more strongly
correlated with accessibility than race itself. Our diversity ratio
measure is large when the amount of segregation in the immediate area
is smaller than the amount of segregation in the greater surrounding
region. In Georgia, this measure tends to be largest in small towns.
When this measure has a constant effect on the overall population's
accessibility, the coefficient takes on a negative value, which implies
that relatively diverse locations experience smaller travel costs.

\begin{figure}
\centering
\begin{tabular}{@{}cc@{}}

\includegraphics{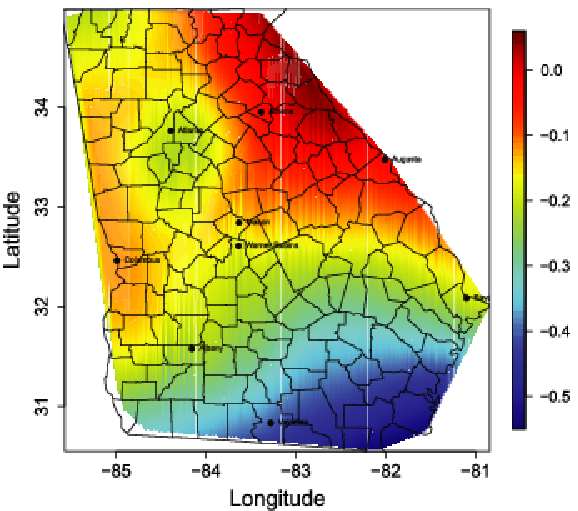}
 & \includegraphics{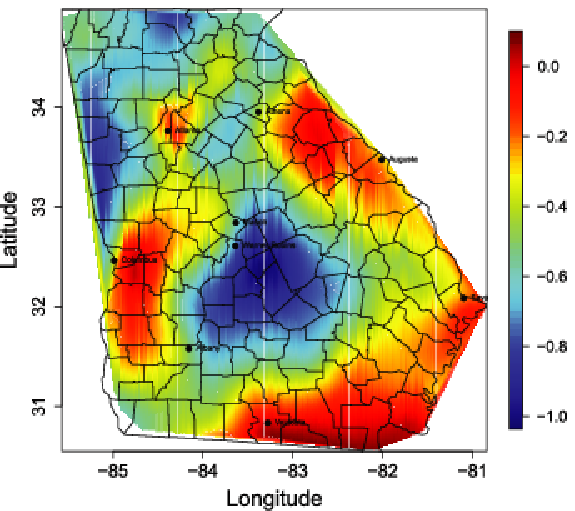}\\
\footnotesize{(a) Percent of population with higher education} & \footnotesize{(b) Population density measurement}\\[3pt]

\includegraphics{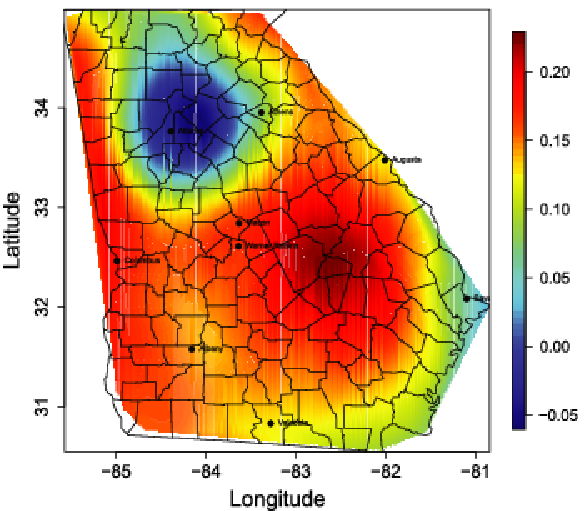}
 & \includegraphics{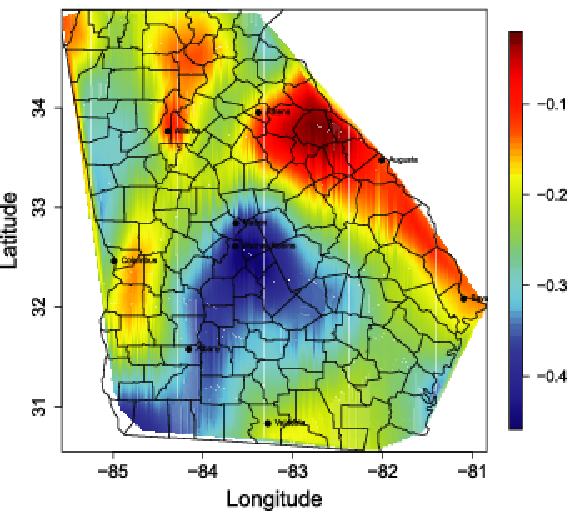}\\
\footnotesize{(c) Distance to hospitals measurement} & \footnotesize{(d) Diversity ratio measurement}
\end{tabular}
\caption{Association map between the listed explanatory factor and the
overall population's travel cost to primary care pediatricians in Georgia.}
\label{fig::non_constant_variables}
\end{figure}

Several variables have nonconstant effects on the overall population's
accessibility in some of the estimated models. Population density has a
consistent nonconstant effect, while our measures of the education
level, distance to hospitals and diversity ratio have constant effects
in some models and nonconstant effects in others. Figure~\ref
{fig::non_constant_variables} shows a sample of these variables'
nonconstant association maps.

%f4 #&#

%{\includegraphics[width=60mm]{plots/NonConstant_Education.pdf}}
%{\includegraphics[width=60mm]{plots/NonConstant_Density.pdf}}
%
%{\includegraphics[width=60mm]{plots/NonConstant_Distance.pdf}}
%{\includegraphics[width=60mm]{plots/NonConstant_Segregation.pdf}}
%

While the inconsistency of these variables' coefficients make
interpretation more difficult, when these variables have a constant
effect on travel cost, the direction of association is as expected:
regions in Georgia with higher education levels, higher population
density and smaller distances to hospitals travel shorter distances to
primary care pediatricians. The median household income has a
significant and positive constant association with accessibility, but
our selection criteria are often superior in models where income is not
included.

The consistency, significance and shape of the coefficients are
identical in models of the overall population's travel cost and in
models of Medicaid patients' travel cost. The range of the constant
coefficients' values are also very similar in the two sets of models.
More detailed results are provided in Supplementary Material D [\citet{supp}]. The consistency of these results is striking,
and implies that the characteristics of vulnerable populations may not
vary with insurance status.

In all of our models there is some variation in accessibility that
remains unexplained. For the overall and Medicaid populations, the
amount of spatial correlation in the models' residuals ranges from
$16.2\%$ to $42.7\%$, and from $34.3\%$ to $43.7\%$, respectively.
There are several plausible explanations. First, researchers may have
overlooked other factors that are the true drivers of accessibility.
Alternatively, these factors may not identify locations where
accessibility is superior to that of other regions with similar
demographics, risk factors and resources. These anomalies are often the
result of local community interventions. The fact that there is more
unexplained variation in models of the Medicaid population's
accessibility suggests that local interventions which focus on
improving Medicaid patients' accessibility may be effective in changing
the odds for these patients. While the concept of ``positive deviance''
has been explored in health outcomes [\citet{Pea02}; Walker et al.
(\citeyear{Waletal})], there is little work on this with respect to accessibility and
exploring this hypothesis would be an interesting direction for future research.

%s5 #&#
\section{Conclusions}\label{sec:summary}

% Summarize what the paper is about.
This paper introduces a comprehensive approach for making inferences
about disparities in spatial accessibility. We develop and implement
methodology for modeling accessibility that accounts for various
constraints in the delivery system, including physicians'
characteristics and capacity. We simultaneously estimate multiple
measures of accessibility including congestion, travel distance and
coverage. By using an optimization-based approach, we can evaluate the
implications of changes in the system, like those caused by policies
which affect physician participation in Medicaid. Our measurement
procedure is general, applicable to different types of care and
scalable to varying geographic domains (e.g., state vs. national) and
different network densities.

% Summarize important findings
Our focus is on pediatric primary care accessibility. Using the models
introduced in this paper, we find that there is a strong association
between a community's coverage rate and travel cost, but that there is
more variability in congestion. The healthcare system is sensitive to
reductions in physicians' Medicaid caseload capacity, but resistant to
many policies designed to improve accessibility. Population density,
distance to hospitals, education and segregation levels are the factors
most strongly associated with patients' travel costs in Georgia.

% Limitations
One limitation of our optimization models is that the assignment
solution is not unique. For example, we found five different ways to
assign children to pediatricians in Georgia and satisfy our models'
constraints. Furthermore, our search was not exhaustive and there may
be many more solutions. However, if two alternative solutions make
small adjustments or trade-offs between immediate neighbors, the
average accessibility in the community will not depend on the
particular solution which is used. Indeed, we compared different
solutions derived from our optimization model using statistical
hypothesis testing, and we found that the difference between solutions
is not statistically significant. We therefore conclude that our
results will not be affected by the solution that we choose.

The precision of our results is also limited by the available data.
Ideally, when implementing our models we would have data on each
physician's caseload and the extent of his or her participation in
Medicaid programs. However, due to the sensitive nature of medical
records and data, we only have aggregate estimates of this information.
More detailed data on patient behavior, including their tolerance for
congestion and the mobility of patients without access to cars, would
also eliminate the need for several additional assumptions in our
model. Finally, because we do not consider physicians located in
neighboring states near the Georgia border, our results may suffer from
edge effects. Therefore, our estimates of accessibility may be too low
in census tracts close to the state line.

As mentioned throughout the paper, there are many avenues for future
research related to accessibility measurement and inference.
Interesting work could be done to model the effects of public
transportation on patients' accessibility, especially in urban areas.
It is also important to consider the impact on accessibility of
policies that would change the structure of the provider network.
Furthermore, in this paper, we do not explore policies' impact on
health outcomes, like the number of emergency department visits.
Addressing this question and further determining the relationship
between accessibility to healthcare and health outcomes is an important
extension of this work. Methods capable of evaluating the association
between accessibility and a very large number of factors, including
those not already highlighted by the literature, may also improve upon
our regression models. Finally, studies should be done to consider
accessibility in different states, for different types of healthcare
and for many additional population groups. While this paper provides a
basis for analyzing patients' accessibility to healthcare, there is
still much work to be done.


\begin{supplement}[id=suppA]
\stitle{Supplement to ``Spatial accessibility of pediatric primary healthcare:
Measurement and inference''}
\slink[doi]{10.1214/14-AOAS728SUPP} %[doi,text={...}] - jei reikia suskaldyti doi
\sdatatype{.pdf}
\sfilename{aoas728\_supp.pdf}
\sdescription{Supplementary Materials A, B, C and D contain four sections [\citet{supp}].
In Supplementary Material A we describe
methods that we utilized in our study but which are not essential
components of our measurement and inference approach. In Supplementary
Material B we give further details about the estimation of our
space-varying coefficient model. In Supplementary Material C we provide
additional details on the data sources we used to implement our models.
In Supplementary Material D we present further results on children's
accessibility to primary care pediatricians in Georgia.}
\end{supplement}

% imsref loaded by akundreckaite, 2014-10-02 16:11:56
%

\printaddresses
\end{document}